\documentclass[journal=langd5,manuscript=article]{achemso}
\usepackage{amsmath}
\usepackage{amssymb}
\usepackage[usenames,dvipsnames]{color}
\usepackage{subfig}
\usepackage{pgf}
\usepackage{mathtools}
\usepackage{gensymb}

\usepackage[version=3]{mhchem} 
\usepackage[T1]{fontenc}

\SectionNumbersOn

\newcommand{\evalAt}[1]{\bigg |_{#1}}

\def\i{\mathbf{i}}
\def\j{\mathbf{j}}
\def\e{\varepsilon}

\author{C. Chalmers, R. Smith and A.J. Archer}
\affiliation{Department of Mathematical Sciences, Loughborough University, Loughborough LE11 3TU, UK}
\email{A.J.Archer@lboro.ac.uk}

\title[Dynamical density functional theory for the evaporation of droplets
  of nanoparticle suspension]
  {Dynamical density functional theory for the evaporation of droplets
  of nanoparticle suspension}

\abbreviations{DFT,DDFT,MC}
\begin{document}

\begin{abstract}

We develop a lattice gas model for the drying of droplets of a nanoparticle suspension on a planar surface, using dynamical density functional theory (DDFT) to describe the time evolution of the solvent and nanoparticle density profiles. The DDFT assumes a diffusive dynamics but does not include the advective hydrodynamics of the solvent, so the model is relevant to highly viscous or near to equilibrium systems. Nonetheless, we see an equivalent of the coffee-ring stain effect, but in the present model it occurs for thermodynamic rather the fluid-mechanical reasons. The model incorporates the effect of phase separation and vertical density variations within the droplet and the consequence of these on the nanoparticle deposition pattern on the surface. We show how to include the effect of slip or no-slip at the surface and how this is related to the receding contact angle. We also determine how the equilibrium contact angle depends on the microscopic interaction parameters.

\end{abstract}

\section{Introduction}\label{introduction}

The structures formed on surfaces from the drying of liquid films or droplets containing suspended colloids or nanoparticles can vary significantly, depending on the nature of the suspended particles, the solvent, the surrounding airflow, the vapour pressure and the nature of the surface \cite{blunt2010patterns, han2012learning, routh2013drying, thiele2014patterned}. Whether the surface is rough or smooth, solvophobic or solvophilic, patterned, curved or in any other way heterogeneous, makes a crucial difference. Understanding the drying dynamics and pattern formation in such systems is not only fascinating fundamental science, but there are many practical application that rely on the behaviour of such fluids at interfaces, ranging from lubrication to the use of ink-jet printing in advanced manufacturing -- see e.g.\ the recent example in Ref.~\citenum{walls}.

Perhaps the most typical example of this is the coffee ring stain formed when a spilt droplet of coffee (or indeed many other liquids containing solutes or suspended particles) dries on a surface \cite{han2012learning, routh2013drying, thiele2014patterned, deegan1997,  larson2014}. As the water in the droplet evaporates, the majority of the coffee is deposited around the edge, in the vicinity of the droplet contact line. This is despite more of the coffee having been initially in the centre, since it is uniformly dispersed within the liquid and the centre is where the liquid thickness above the surface is greatest. The coffee stain is formed because the evaporation of the liquid leads to a hydrodynamic flow of liquid from the centre of the droplet towards the edge. This flow carries the suspended particles to the edge of the droplet, where they remain when all the liquid is finally evaporated \cite{routh2013drying, thiele2014patterned, deegan1997, larson2014}. Note however that ring deposition does not alway occur. For example, it can be suppressed if the particles are elongated instead of roughly spherical \cite{yunker2011suppression}. But in general the effect must be overcome in applications requiring uniform surface deposition, such as in printing and coating.

Modelling such intricacies presents a challenge, because they depend on a fine balance and interplay of thermodynamic effects related to evaporation and perhaps also phase transitions within the droplet and hydrodynamic effects related to fluid flow within the droplet and overall droplet dynamics. These considerations are especially important if there are advancing or receding contact lines \cite{bonn2009wetting}. Thin film hydrodynamic models have been used to to describe many key aspects \cite{oron1997, kalliadasis2007, thiele2009, dietrich, frastia2011, frastia2012}, but such models are unable to describe vertical variations in the local particle concentration and do not fully capture any of the microscopic structure within the liquid. { Many of these issues are addressed by using lattice Boltzmann (LB) methods,\cite{kruger2017lattice} which model the solvent as a continuum, but the suspended particles are treated explicitly -- see for example the LB models for the evaporation of droplets containing suspended particles described in Refs.\ \citenum{joshi2010wetting, zhao2017modeling}.} Fully microscopic models based on molecular dynamics (MD) computer simulations, such as those described in Refs.\ \citenum{ingebrigtsen2007, lu2013, tretyakov2013, becker2014}, do of course describe every aspect of the structure and dynamics in the droplet. However, these approaches are limited in the size of droplet that can be modelled. The same is to some extent also true when classical density functional theory (DFT)~\cite{Evans79, Evans92, hansen} and dynamical DFT (DDFT)~\cite{MaTa99, MaTa00, ArEv04, ArRa04, archer06, archer09} are applied to describe the structure and dynamics of droplets -- see e.g.\ Refs.\ \citenum{hughes2014, nold2014, nold2015nanoscale, hughes2015, hughes2017} and references therein. That said, because DFT and DDFT are statistical mechanical theories, the scaling of the computational cost with size of the system is better than the scaling with MD \cite{goddard2012unification}.

Here we develop a DDFT for droplets containing suspended nanoparticles that is based on a lattice Hamiltonian (generalised Ising) model of the microscopic interactions in the system. The advantage of assuming the nanoparticle suspension can be modelled as a lattice fluid is that it allows us to describe much larger droplets than is feasible using a fully microscopic DFT -- see e.g.\ Ref.~\citenum{hughes2017}.

Lattice models have been successfully used previously to describe the evolution of liquids and particle suspensions on surfaces. These models were initially 2-dimensional (2D) Monte Carlo (MC) models \cite{blunt2010patterns, rabani2003, pauliac2008, vancea08, stannard2011dewetting, vancea2011pattern} that assumed there is no vertical variation in the liquid in the direction perpendicular to the solid surface on which the liquid is deposited. However, 3-dimensional (3D) MC models have subsequently also been developed \cite{sztrum2005, kim2011, jung2014, crivoi2014, tewes2017comparing, chalmers2017}. The present DDFT assumes the same Hamiltonian as the MC model in Ref.~\citenum{chalmers2017}. Thus, the DDFT presented here is able to fully describe any vertical variations in the local densities within the droplet, such as the formation of a nanoparticle `crust' on the drying droplet, unlike the 2D DDFT models developed previously \cite{archer2010dynamical, robbins2011}.

An important aspect of the MC model in Ref.~\citenum{chalmers2017} is the identification of how the interactions between lattice sites should vary with distance in order to have liquid droplets with a realistic hemispherical shape, so as to lessen the influence of the underlying grid. Owing to the fact that we base our DDFT on the same Hamiltonian, the present model also has this advantage. An additional feature of the DDFT developed here is that it incorporates the effects of slip, no-slip or partial slip of the liquid at the surface. We show how this affects the evolution of the shape of droplets as they evaporate and how this is connected to the receding contact angle. Since the DDFT incorporates all the thermodynamics related to the degree of solubility of the nanoparticles in the solvent liquid, the model incorporates the effects of phase separation (aggregation) of the nanoparticles as the local nanoparticle concentration increases due to the solvent evaporation. We show that this can lead to a coffee-ring like stain. However, in the present model it is due to the thermodynamics of phase separation, not the usual advective fluid mechanical mechanism \cite{routh2013drying, thiele2014patterned, deegan1997, larson2014}. The DDFT we use is one that assumes only diffusive particle motion -- i.e.\ we assume the droplet is not too far from equilibrium. This is the original DDFT of Refs.\ \citenum{MaTa99, MaTa00, ArEv04, ArRa04} for both the solvent and the suspended particles, rather than more sophisticated DDFTs that include inertial effects \cite{archer06, archer09} or effects of hydrodynamics \cite{rex2009dynamical, goddard2012unification, goddard2012general}. We also show that the thermodynamics of phase separation can lead to the deposition of multiple rings.

{ This paper is structured as follows: In Section~\ref{methodology} we introduce the lattice Hamiltonian for the system and the approximation we use for the free energy of the system that is the input to the DDFT. In Section~\ref{bulk-phase} we describe the bulk-mixture phase behaviour and present phase diagrams showing how the vapour-liquid phase separation depends on the model interaction parameters and changes as the concentration of the nanoparticles is varied. In Section~\ref{equilibrium} we calculate equilibrium interfacial properties, including calculating the density profiles of the solvent and the nanoparticles at the vapour-liquid interface, the surface tension and how it depends on temperature and also the equilibrium contact angle. We also compare with the MC results from Ref.\ \citenum{chalmers2017} to illustrate the accuracy of the DFT. In Section~\ref{dynamics} we describe the DDFT used to describe the non-equilibrium dynamics of the solvent and the nanoparticles, including how to include the effects of (no)slip at the substrate. In Section~\ref{evaporation} we display the results from simulating evaporating droplets of both pure solvent and also containing nanoparticles. We show how the receding contact angle depends on the parameters controlling the slip at the surface and present results for the deposits left by evaporating droplets including a coffee-ring like deposit, a deposit equivalent to multiple rings and also patterns related to spinodal dewetting. Finally, in Section~\ref{conclusion} we make a few concluding remarks.}

\section{Lattice model for the system}\label{methodology}

We model the nanoparticle suspension by discretising onto a 3D cubic lattice with lattice spacing $\sigma$ (which is also the diameter of the particles), with each site on the lattice labelled by the index $\i$, where $\i=(i,j,k)\in \mathbb{Z}^3$. Thus, $\i$ defines the location of the lattice site. Henceforth, we set $\sigma = 1$, defining our unit of length. The energy of the system (Hamiltonian) is given by
\begin{equation}
  \begin{aligned}
  E =& - \sum_{\i,\j} \left(
      \frac12 \e^{ll}_{\i\j}n^l_{\i}n^l_{\j}
    + \e^{nl}_{\i\j} n^l_{\i}n^n_{\j}
    + \frac12 \e^{nn}_{\i\j} n^n_{\i}n^n_{\j}
  \right)\\
 &\quad - \mu\sum_{\i} n^l_{\i}
  + \sum_{\i} \Phi_{\i}^l n^l_{\i}
  + \sum_{\i} \Phi_{\i}^n n^n_{\i},
\end{aligned}
\label{eq:mc-hamiltonian}
\end{equation}
where $n^l_\i$ is the occupation number for the liquid at site $\i$ and
$n^n_\i$ is the occupation number for nanoparticles at site $\i$,
i.e., $n^l_\i = 1$ if the site is occupied by liquid and $n^l_\i=0$ if
unoccupied by liquid. Similarly, $n^n_\i = 0$ or $1$ depending on
whether or not the site $\i$ is occupied by a nanoparticle. A lattice site
cannot be occupied by both liquid and a nanoparticle. {Note
that treating the liquid in this manner corresponds to a
coarse-grained treatment, since each lattice site occupied by the
liquid does not
represent one individual solvent molecule, but a group of molecules
occupying the volume $\sigma^3$. Since there are no significant qualitative differences between results from 2D MC models where the diameter of the nanoparticles is set to be three times that of the liquid `particles' and results from models where the diameters are equal,\cite{blunt2010patterns, rabani2003, pauliac2008, vancea08, stannard2011dewetting, vancea2011pattern} this justifies setting the lattice spacing equal to the diameter of the nanoparticles. Indeed, one can even argue that $\sigma$ is greater than the diameter of the nanoparticles and that just as a site `occupied' by liquid corresponds to `mostly containing liquid', a site `occupied by nanoparticles' corresponds to `mostly containing nanoparticles', i.e.\ $\sigma$ can be viewed as a coarse-graining length-scale.\cite{chalmers2017}}

{In Eq.~\eqref{eq:mc-hamiltonian}} $\Phi^l_\i$ is the external potential due
to the surface influencing the liquid at site $\i$  and $\Phi^n_\i$ is
the external potential for the nanoparticles. The interaction between
pairs of liquid particles at sites $\i$ and $\j$ is determined by the
matrix $\e^{ll}_{\i\j} = \e_{ll}c_{\i\j}$, which is a discretised pair
potential. The parameter $\e_{ll}$ governs the overall strength.
Similarly, $\e^{ln}_{\i\j} = \e_{ln}c_{\i\j}$ is the interaction matrix
between nanoparticles and liquid, with strength determined by the
parameter $\e_{ln}$, and $\e^{nn}_{\i\j} = \e_{nn}c_{\i\j}$ is the
interaction between pairs of nanoparticles, with $\e_{nn}$ determining
the overall strength. $c_{\i\j}$ is a dimensionless coefficient which
decreases in value as the distance between the pairs of particles
increases. We use the following values
\begin{equation}
  c_{\i\j} =
  \begin{dcases}
    1            & \j \in \{\text{NN } \i\}   \\
    \frac{3}{10} & \j \in \{\text{NNN } \i\}  \\
    \frac{1}{20} & \j \in \{\text{NNNN } \i\} \\
    0            & \text{otherwise}
  \end{dcases}
\label{eq:c_ij}
\end{equation}
where NN $\i$, NNN $\i$ and NNNN $\i$ stand for nearest neighbours, next
nearest neighbours and next-next nearest neighbours, respectively, to lattice site $\i$. The
choice of particular values in Eq.~\eqref{eq:c_ij} is important, as this
leads to liquid droplets on the surface having a hemispherical
shape \cite{chalmers2017}. For example, if instead we just have nearest
neighbour interactions, then the system forms unrealistic shaped
droplets with facetted surfaces, particularly at low temperatures.

If the nanoparticle suspension is in contact with a planar solid surface, this exerts external potentials that we assume are
\begin{equation}
  \Phi^l_\i=
    \begin{cases*}
      -\frac{12}{5}\e_{wl} & $j$ = 0,\\
      0 & otherwise,
    \end{cases*}\\
\label{eq:phi-l}
\end{equation}
and
\begin{equation}
  \Phi^n_\i=
    \begin{cases*}
      -\frac{12}{5}\e_{wn} & $j$ = 0,\\
      0 & otherwise,
    \end{cases*}
\label{eq:phi-n}
\end{equation}
where $\e_{wl}$ is the attraction between the surface and the liquid and
$\e_{wn}$ is the attraction between the surface and the nanoparticles,
and $j$ is the component of $\i$ that varies in the direction
perpendicular to the surface. The factor $12/5$ in
Eq.~\eqref{eq:phi-l} comes from assuming a pair potential
$\e^{wl}_{\i\j} = \e_{wl} c_{\i\j}$ between wall lattice sites and
the liquid. When a liquid particle is next to the wall, summing over
the interaction with the neighbouring wall particles leads
to Eq.~\eqref{eq:phi-l}. Similarly, summing over
$\e^{wn}_{\i\j} = \e_{wn} c_{\i\j}$ leads to
Eq.~\eqref{eq:phi-n}.

For the lattice model defined above, one can study both the equilibrium
and non-equilibrium behaviour using the Monte-Carlo simulation approach
developed in {Ref.}~\citenum{chalmers2017}. Here, a statistical mechanical
theory based on DFT \cite{Evans79, Evans92, hansen} and
DDFT \cite{MaTa99, MaTa00, ArEv04, ArRa04} is derived. Thus,
we develop a theory for the average densities,
\begin{equation}
  \rho^l_\i = \langle n^l_\i \rangle \quad \text{and} \quad
  \rho^n_\i = \langle n^n_\i \rangle,
\end{equation}
which are the ensemble average densities at site $\i$, i.e.,
$\langle\ldots\rangle$ denotes a statistical average. Making a mean
field approximation, the Helmholtz free energy for the binary
lattice-gas is \cite{hughes2014, woywod03, woywod04, woywod06, hughes2015}
%
\begin{align}
F(\{\rho^l_\i\},\{\rho^n_\i\})%
  &= k_B T \sum_\i \left[
    \rho^l_\i \ln \rho^l_\i
    - (1 - \rho^l_\i - \rho^n_\i) \ln (1 - \rho^l_\i - \rho^n_\i)
    + \rho^n_\i \ln \rho^n_\i
  \right]
  \notag\\
  &\quad
    - \frac12 \sum_{\i,\j} \e^{ll}_{\i\j} \rho^l_\i \rho^l_\j
    - \sum_{\i,\j} \e^{ln}_{\i\j} \rho^l_\i \rho^n_\j
    - \frac12 \sum_{\i,\j} \e^{nn}_{\i\j} \rho^n_\i \rho^n_\j
  \notag\\
  &\quad
    + \sum_\i \left(
        \Phi^l_\i \rho^l_\i
      + \Phi^n_\i \rho^n_\i
    \right),
\label{eq:helmholtz}
\end{align}
%
where $k_B$ is Boltzmann's constant and $T$ is the temperature. The above is a discretised DFT {free energy functional} for a binary mixture.

\section{Bulk solvent phase behaviour}\label{bulk-phase}

The densities can be constants when $\Phi^l_\i = \Phi^n_\i = 0$, i.e., we
have a uniform fluid with $\rho^l_\i = \rho_l$ and $\rho^n_\i = \rho_n$, for
all $\i$. In this case, the sum over neighbours in the interaction terms in
the Helmholtz free energy~\eqref{eq:helmholtz} can be evaluated. The
integrated interaction matrix is $\sum_\j c_{\i\j} = 10$ for all $\i$, so we
have $a_{ll} = 10\e_{ll}$, $a_{ln}=10\e_{nl}$ and $a_{nn}=10\e_{nn}$ as
the integrated strengths of the pair interactions. From
Eq.~\eqref{eq:helmholtz} the average Helmholtz free energy per lattice
site, $f = F/V$, where $V$ is the volume of the system, is given by
\begin{align}
  f &=
  k_BT \left(
     \rho_l \ln \rho_l
     + (1 - \rho_l - \rho_n) \ln (1 - \rho_l - \rho_n)
     + \rho_n \ln \rho_n
   \right)
\notag\\
   &\qquad - \frac12 a_{ll} \rho_l^2
   - a_{ln} \rho_l \rho_n
   - \frac12 a_{nn} \rho_n^2.
\label{eq:area-energy}
\end{align}

From this we can calculate the spinodal, the locus
where $\partial^2f/\partial\rho^2 = 0$ and where
$\rho = \rho_l + \rho_n$ is the total density. The spinodal defines the
boundary of the region of the phase diagram where
the system is unstable and density fluctuations in a uniform system
spontaneously grow, leading to phase separation.
For temperatures where two-phase coexistence can occur, the binodal
curve gives the coexisting density values for a system in equilibrium.
This is calculated by equating the chemical potential, temperature and
pressure in each of the coexisting phases. States in the phase diagram
outside the binodal are stable and no phase separation occurs.

We can use Eq.~\eqref{eq:area-energy} to calculate the binodal since
thermodynamic quantities such as the chemical potentials, $\mu_l$ and
$\mu_n$, and pressure, $P$, may be obtained using the following relations
\begin{equation}
  \mu_l = \frac{\partial f}{\partial \rho_l},
  \quad \mu_n = \frac{\partial f}{\partial \rho_n},
  \quad P = -\frac{\partial f}{\partial V}.
\end{equation}
These give
\begin{align}
  \mu_l &= k_BT (\ln \rho_l - \ln(1 - \rho_l - \rho_n))
         - a_{ll} \rho_l
         - a_{ln} \rho_n,
\\
  \mu_n &= k_BT (\ln \rho_n - \ln(1 - \rho_l - \rho_n))
         - a_{ln} \rho_l
         - a_{nn} \rho_n,
\\
  P  &= -k_BT \ln(1 - \rho_l - \rho_n)
       - \frac12 a_{ll} \rho_l^2
       - a_{ln} \rho_l \rho_n
       - \frac12 a_{nn} \rho_n^2,
\label{eq:quantities}
\end{align}
where we have used the fact that in a uniform system the densities
$\rho_l = N_l/V$ and $\rho_n = N_n/V$, where $N_l$ and $N_n$ are the
total numbers of each species in the system.

\begin{figure}
\centering
  \includegraphics[width=1.\columnwidth]{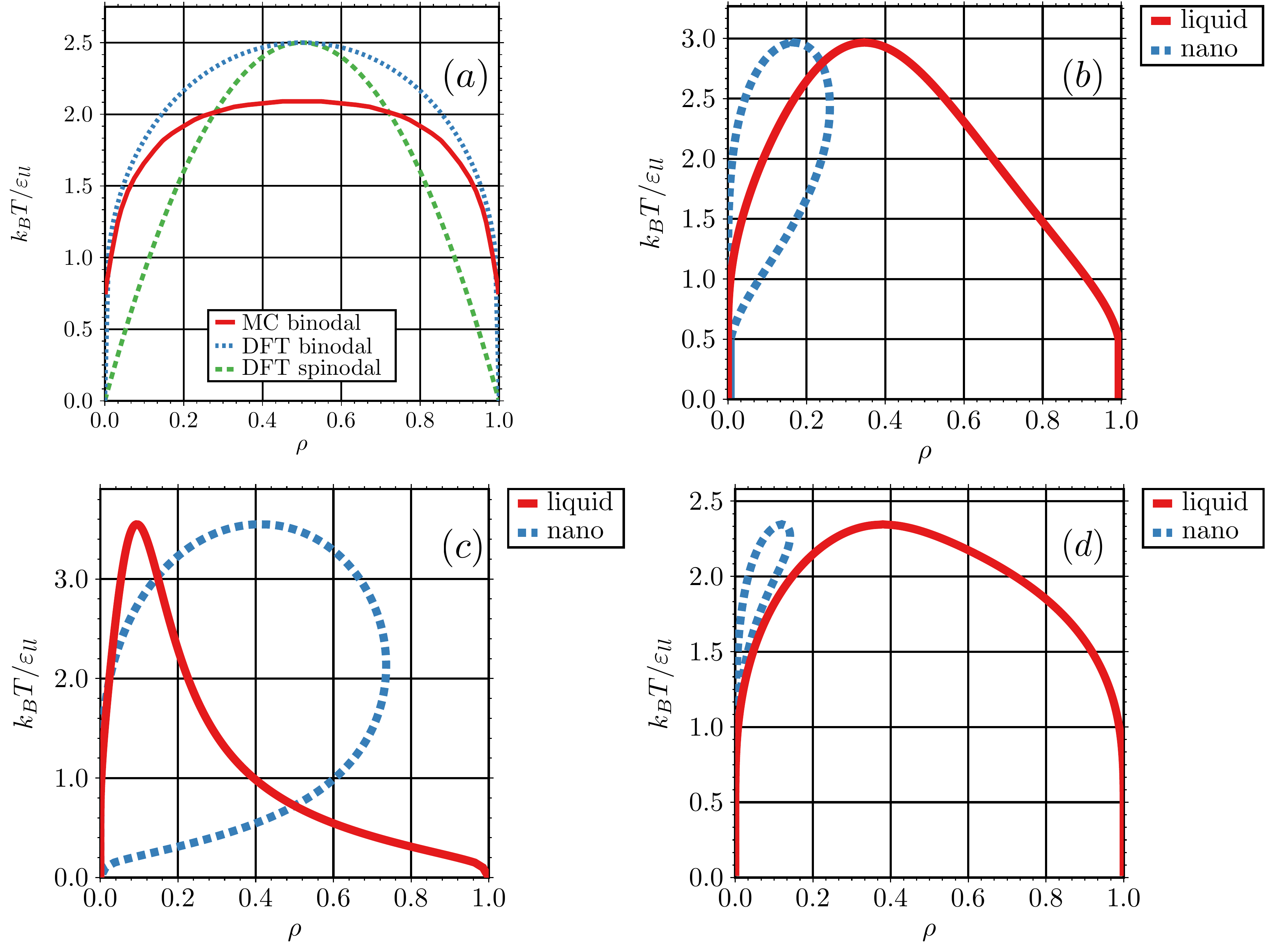}
\caption{%
  {Binodal curves for the fluid in the temperature
  versus density plane. In (a) we display the results for the pure
  liquid ($\mu_n=-\infty$) from
  both DFT and MC computer simulations and also
  display the spinodal from DFT. The MC
  results are from Ref.~\citenum{chalmers2017}.
  In (b) we display the binodal curves for the binary mixture
  with $\mu_n/\e_{ll}=-10$,
  $\e_{nl}/\e_{ll}=1.25$, $\e_{nn}/\e_{ll}=1.5$ and
  in (c) for $\mu_n/\e_{ll}=-8$. In 
  (d) we display the binodal curves for the binary mixture with
  $\mu_n/\e_{ll}=-8$, $\e_{nl}/\e_{ll}=0.75$ and $\e_{nn}/\e_{ll}=1.5$,
  corresponding to nanoparticles which are less soluble in the liquid.
  }
}
\label{fig:liquid-binodal}
\end{figure}

For the pure liquid with no nanoparticles present ($\rho_n = 0$), we can
use the symmetry of the Hamiltonian~\eqref{eq:mc-hamiltonian} to
simplify the calculation of the binodal \cite{robbins2011}. This allows
us to observe that if $\rho_l$ is the density of the liquid at
coexistence then $(1-\rho_l)$ is the density of the coexisting vapour.
On equating the pressure in the two phases we obtain the following
equation for the binodal:
\begin{equation}
  \frac{k_BT}{\e_{ll}} = \frac{5(2\rho_l - 1)}{\ln[\rho_l/(1-\rho_l)]}.
\end{equation}
This has a maximum at $\rho_l = 0.5$ which corresponds to a critical
temperature of $k_BT/\e_{ll} = 2.5$. Fig.~\ref{fig:liquid-binodal}{(a)} shows
a plot of this binodal curve together with the binodal from
Ref.~\citenum{chalmers2017} that was calculated using MC simulations for
the same system, with Hamiltonian given by Eqs.\ \eqref{eq:mc-hamiltonian} and \eqref{eq:c_ij}. The binodals are qualitatively similar, but at higher temperatures there is a sizeable difference between the two curves since
the DFT in Eq.~\eqref{eq:helmholtz} is a mean field theory. The critical
temperature predicted by the DFT is around $0.4 k_B/\e_{ll}$ higher than
the true value. However, for temperatures $k_BT/\e_{ll} < 1.5$ we see
that the coexisting density values from the DFT are in fairly good
agreement with those from the MC. This is the regime in which the results below are obtained.

Returning to consider the full binary mixture, to calculate the
binodals we have the additional condition that the chemical
potential of the nanoparticles, $\mu_n$, is the same in both phases. The
phase diagram is no longer symmetric around $\rho_l=0.5$ when the
nanoparticles are present. Thus, we must solve the following
simultaneous equations:
\begin{align}
  T^{\text{LDP}} &= T^{\text{HDP}}, \\
  P^{\text{LDP}} &= P^{\text{HDP}}, \\
  \mu_l^{\text{LDP}} &= \mu_l^{\text{HDP}}, \\
  \mu_n^{\text{LDP}} &= \mu_n^{\text{HDP}},
\end{align}
where LDP stands for low density phase, HDP stands for high density
phase. The first equation can be trivially satisfied by simply setting
the same temperature in both phases. We then also fix the chemical
potential of the nanoparticles to some specified value. This then gives
us four equations for the four unknowns, namely the densities of the two
species in the two different phases~\cite{robbins2011}. Solving like
this for a range of temperatures gives us the phase diagram.

{Fig.~\ref{fig:liquid-binodal}(b)} and {(c)} show the binodals for the liquid-nanoparticle mixture for the case when $\e_{nl}/\e_{ll} = 1.25$ and $\e_{nn}/\e_{ll} = 1.5$ and for different values of the nanoparticle chemical potential $\mu_n$. We see that as $\mu_n$ is increased the density of the nanoparticles increases in both phases and can in fact become the majority species for large enough $\mu_n$. Note that Fig.~\ref{fig:liquid-binodal}{(a)} can be considered to be the $\mu_n =-\infty$ case in this sequence with varying $\mu_n$, where, of course, $\rho_n = 0$ in both coexisting phases.

In Fig.~{\ref{fig:liquid-binodal}(d)} we show results for a case where
$\e_{nl}$ is less than both $\e_{nn}$ and $\e_{ll}$, in contrast to the
case in {panels (b) and (c)}, where
$\e_{nl} = \frac12(\e_{nn} + \e_{ll})$. In this case it is energetically
unfavourable for the nanoparticles to mix with the liquid
and so for the case in {Fig.~\ref{fig:liquid-binodal}(d)} where
$\mu_n/\e_{ll} = -8$ (a low value), the density of the nanoparticles in
both coexisting phases is low. For higher values of $\mu_n$ (not
displayed) we observe liquid-liquid phase separation similar to that
described in Ref.~\citenum{archer13}.


\begin{figure}
\includegraphics[width=0.49\columnwidth]{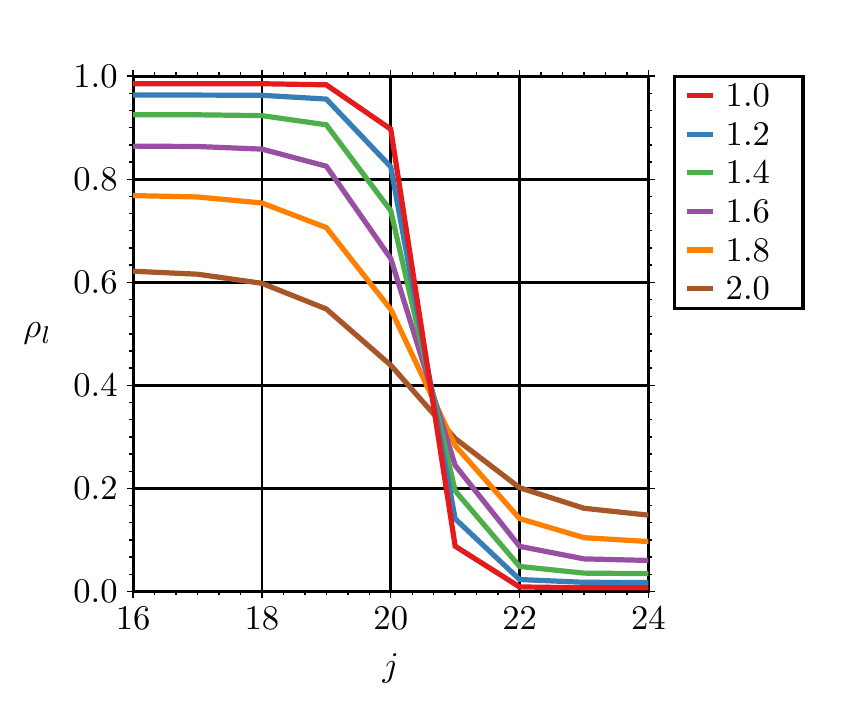}
\includegraphics[width=0.49\columnwidth]{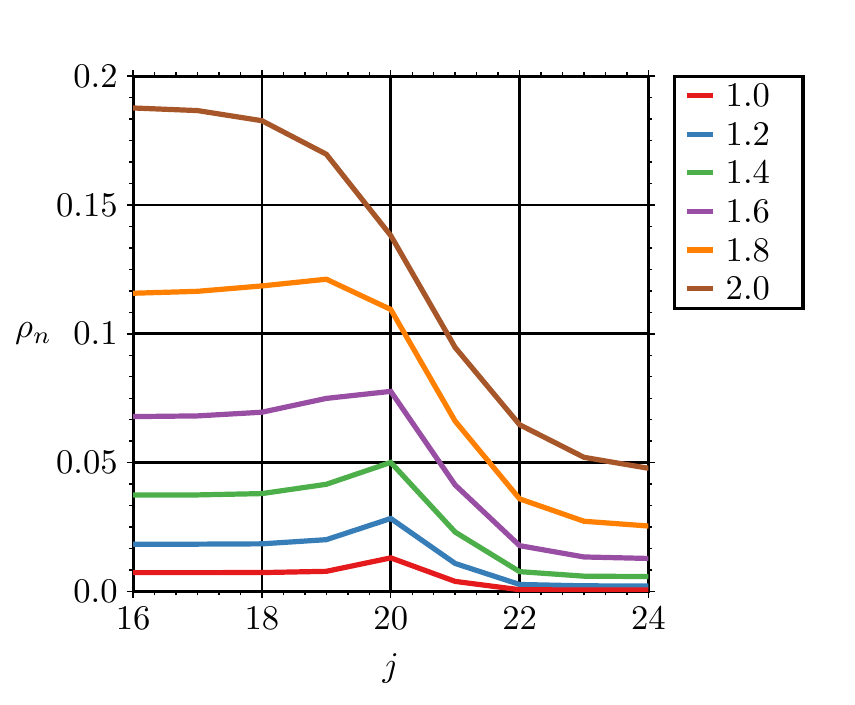}
\caption{%
  Liquid density profiles (left) and nanoparticle density profiles
  (right) at the free liquid-vapour interface for
  $\e_{nl}/\e_{ll}=0.75$, $\e_{nn}/\e_{nl} = 1.5$ at different
  values of $k_BT/\e_{ll}$, the dimensionless temperature,
  as indicated in the key.
  The corresponding bulk fluid phase diagram is in
  Fig.~{\ref{fig:liquid-binodal}(d)}.
}
\label{fig:liquid-equ}
\end{figure}

\section{Equilibrium interfacial behaviour}\label{equilibrium}

Having determined the bulk fluid phase behaviour, we now briefly
consider the interface between the coexisting phases and calculate the
surface tension.

\subsection{Density profiles at the free interface}

At the planar interface between the vapour and the liquid phases the density
profiles vary only in the direction perpendicular to the interface. We
assume that the index varying in the direction perpendicular to the
interface is $j$. Recall $\i=(i,j,k)$. The density profiles are
calculated by minimising the grand potential
\begin{equation}
  \Omega = F - \mu_l \sum_\i \rho^l_\i - \mu_n \sum_\i \rho^n_\i
\label{eq:omega}
\end{equation}
where the Helmholtz free energy $F$ is given by Eq.~\eqref{eq:helmholtz}
and the chemical potentials $\mu_l$ and $\mu_n$ are set to be the values
at which vapour-liquid phase coexistence occurs. In
Fig.~\ref{fig:liquid-equ} we display the density profiles of the solvent
and the nanoparticles for the case when $\mu_n/\e_{ll}=-8$,
$\e_{nl}/\e_{ll}=0.75$ and $\e_{nn}/\e_{ll}=1.5$ and various
temperatures. The corresponding bulk fluid phase diagram is displayed in
Fig.~{\ref{fig:liquid-binodal}(d)}. We see that as the temperature is
increased the total density difference between the two coexisting phases
decreases. We note also that at lower temperatures, $k_BT/\e_{ll}
\lesssim 1.8$, there is a small enhancement of the nanoparticle density
at the interface, indicating that the nanoparticles have a slight
propensity towards being surface active~\cite{thinfilms} for these
values of $\e_{nl}$ and $\e_{nn}$. As we show below, this slight surface
enhancement in equilibrium can become greater during non-equilibrium droplet evaporation.

\subsection{Surface tension}\label{surface-tension}

Having calculated interfacial density profiles such as those in the
Fig.~\ref{fig:liquid-equ}, we can then substitute back into
Eq.~\eqref{eq:omega} to calculate the grand potential, $\Omega$, of the
whole system, including the interface.

The surface tension is defined as the excess free energy in the system
due to the presence of an interface between two coexisting phases.
Subtracting the grand potential $\Omega_0 = -PV$ for a system with the
same volume $V$, temperature and chemical potentials but containing
either only the uniform vapour or liquid gives the excess grand potential due to
the interface. The interfacial tension is then
\begin{equation}
  \gamma = \frac{\Omega + PV}A,
\end{equation}
where $A$ is the area of the interface. From the density profiles in
Fig.~\ref{fig:liquid-equ}, this gives $\gamma_{gl}$, the
planar liquid-gas interfacial tension. In a similar manner, for the
fluid at the wall exerting potentials $\Phi^l_\i$ and $\Phi^n_\i$ we can
calculate $\gamma_{wl}$, the wall-liquid interfacial tension and
$\gamma_{wg}$, the wall-gas interfacial tension. These are all
calculated for $\mu_l = \mu_l^{\text{coex}}$ and $\mu_n =
\mu_n^{\text{coex}}$, the values at bulk gas-liquid phase coexistence.

From these interfacial tensions one can then calculate the contact angle
a droplet of liquid would have with the surface using Young's
equation \cite{de2004capillarity},
\begin{equation}
  \theta = \arccos \left(
    \frac{\gamma_{wg} - \gamma_{wl}}{\gamma_{lg}}
  \right).
\label{eq:youngs}
\end{equation}

\begin{figure}
\centering
  \includegraphics{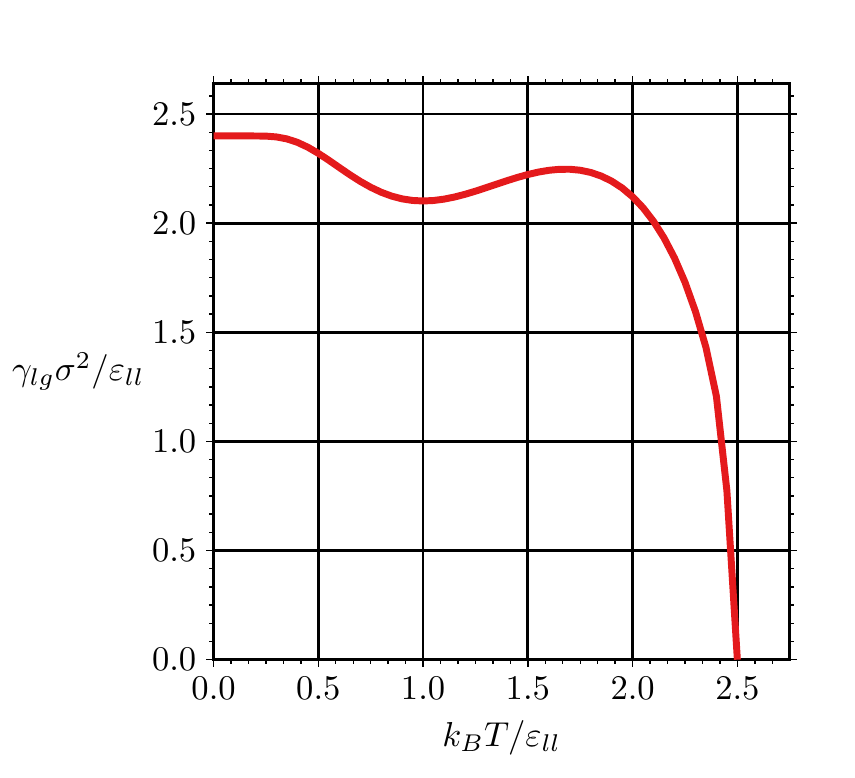}
\caption{%
  The liquid-gas surface tension as a function of temperature,
  calculated using DFT. The corresponding bulk fluid phase diagram is
  in Fig.~\ref{fig:liquid-binodal}.
}
\label{fig:surface-tension}
\end{figure}

Fig.~\ref{fig:surface-tension} shows the liquid-gas surface tension for
the pure liquid ($\mu_n = -\infty$) plotted as a function of temperature.
There is a slight local minimum in the surface tension at $k_BT/\e_{ll}\approx
1.0$. In the limit $T \rightarrow 0$ the density of the liquid $\rho_l
\rightarrow 1$ and the coexisting gas has $\rho_l \rightarrow 0$. It is
then straightforward to see from Eq.~\eqref{eq:mc-hamiltonian} or
Eq.~\eqref{eq:helmholtz} that for $T \rightarrow 0$ the surface tension
$\gamma_{lg}/\e_{ll} \rightarrow 12/5$. At the critical temperature, $T
= T_c = 2.5 \e_{ll}/k_B$, the density difference between the two
coexisting phases goes to zero, so as $T \rightarrow T_c$, $\gamma_{lg}
\rightarrow 0$.

\begin{figure}
\centering
  \includegraphics{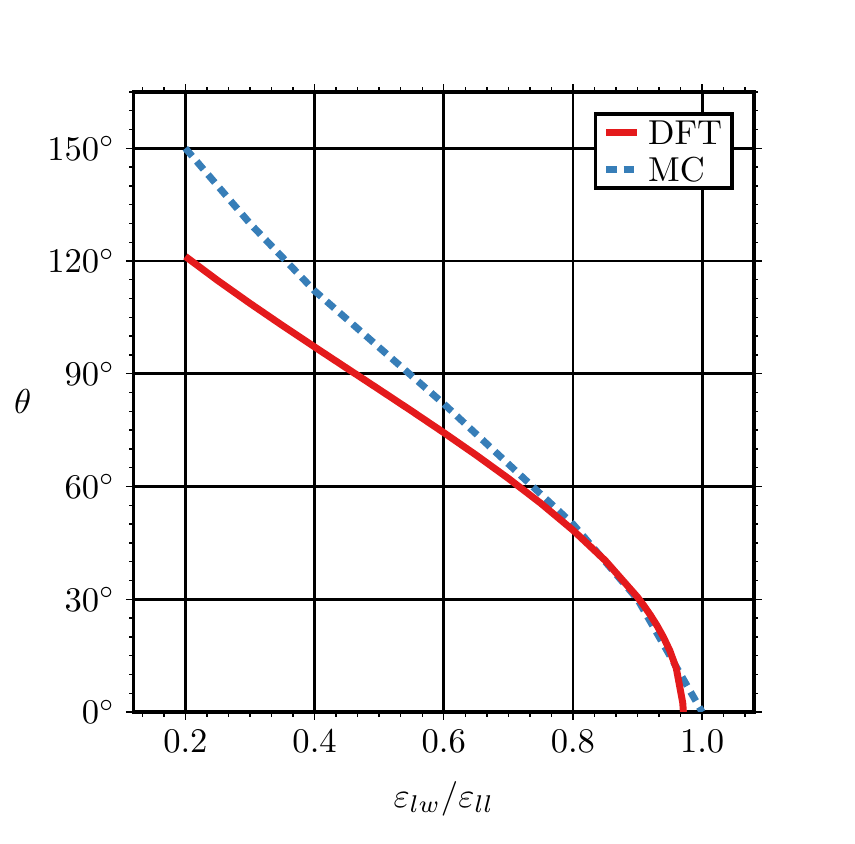}
\caption{%
  The contact angle for a droplet on a surface with temperature
  $k_BT/\e_{ll} = 1.0$. The solid line is the result from DFT using the
  interfacial tensions together with Young's equation~\eqref{eq:youngs}.
  The dashed line is the corresponding results from MC simulations \cite{chalmers2017}.
}
\label{fig:contact-angle}
\end{figure}

Fig.~\ref{fig:contact-angle} shows the contact angles calculated from the
DFT for the pure liquid at the temperature $k_BT/\e_{ll} =
1.0$, as the attraction due the surface is varied. We see that for $\e_{lw}/\e_{ll}>0.97$ the contact angle $\theta=0$, i.e.\ the liquid wets the surface. In contrast, for $\e_{lw}/\e_{ll}<0.97$ the liquid is only partially wetting. As the attraction to the surface $\e_{lw}$ is decreased below this value, the contact angle $\theta$ increases and can become large, as the surface becomes increasingly solvophobic. It is also possible to directly obtain the contact angle by
fitting a circle to an equilibrium droplet density profile calculated
using DFT (e.g.\ using the method in Ref.~\citenum{taubin}), which gives almost identical results \cite{hughes2015}.
Such a droplet density profile is calculated by constraining the total
volume of liquid in the system to be fixed and also allowing the density
to vary in both perpendicular and parallel directions to that surface --
see Ref.~\citenum{hughes2015} for further details.
For comparison, in Fig.~\ref{fig:contact-angle} we also show the contact angles obtained by fitting a
circle to equilibrium density profiles calculated in Ref.~\citenum{chalmers2017}
using MC simulations of droplets on surface, using
Hamiltonian~\eqref{eq:mc-hamiltonian}. At higher values of
$\e_{lw}/\e_{ll}$ the curves are almost identical but at lower values
the MC simulations give a higher contact angle. We believe this is due
to the fact that at higher values of $\e_{lw}$ the energetic
contributions to the free energy dominate and so the DFT describes the
droplet accurately. However, for smaller $\e_{lw}$, i.e., a solvophobic
surface, the fluctuations of the liquid near the surface and in the
contact line region are significant \cite{evans2015,chacko2017} and so
the mean-field DFT, which neglects some fluctuation contributions to the
free energy, does less well. We should emphasise, however, that
interfacial tensions and especially the contact angle $\theta$ are
quantities that depend very sensitively on approximations, so the agreement in Fig.\ \ref{fig:contact-angle} is actually rather good.

\section{Theory for the non-equilibrium dynamics}\label{dynamics}

We assume that the non-equilibrium fluid dynamics is described by DDFT~\cite{MaTa99, MaTa00, ArEv04, ArRa04}. This is a good assumption to make for the nanoparticle dynamics, since it is a theory for Brownian particles suspended in a liquid, as derived in Refs.~\citenum{MaTa99, MaTa00, ArEv04, ArRa04}. However as discussed in Refs.~\citenum{archer06, archer09}, the theory can also approximate the dynamics of molecular liquids, especially when the fluid is not too far from equilibrium, which is certainly true for the cases of interest here. For a two component system, DDFT generalises to give the following pair of coupled equations \cite{archer05}
\begin{align}
\frac{\partial \rho^l_\i}{\partial t}
  &= \nabla \cdot
  \left[
    M^l_\i \rho^l_\i \nabla \frac{\partial F}{\partial \rho_{\i}^l}
  \right],
\label{eq:ddftl}
  \\
\frac{\partial \rho^n_\i}{\partial t}
  &= \nabla \cdot
  \left[
    M^n_\i \rho^n_\i \nabla \frac{\partial F}{\partial \rho_{\i}^n}
  \right],
\label{eq:ddftn}
\end{align}
where $M^l_\i$ and $M^n_\i$ are the mobility coefficients for the liquid
and nanoparticles and $F$ is the Helmholtz free energy. The average densities of the liquid and nanoparticles at site $\i$, $\rho^l_\i$ and $\rho^n_\i$, respectively, are now both functions of time $t$. Note
that since the system we consider here is a lattice model, the $\nabla$
operators in Eqs.~\eqref{eq:ddftl} and \eqref{eq:ddftn} are implicitly
the finite difference approximations. For more details on this, see
Ref.~\citenum{robbins2011} and also the Appendix below.

Here, we generalise the mobilities $M^l_\i$ and $M^n_\i$ to be mobility
matrices which depend on both position $\i$ and the direction of the
fluid flow, so as to model the effect of slip, partial-slip or no-slip
at the surface. Thus we set the mobility matrix at site $\i$ for species
$c$ to be
\begin{equation}
  M^c_\i =
  \begin{cases*}
    m^c_\i
    \begin{pmatrix}
      s & 0 & 0 \\
      0 & v & 0 \\
      0 & 0 & s
    \end{pmatrix}
      & $j=1$,\\
    m^c_\i
    \begin{pmatrix}
      1 & 0 & 0 \\
      0 & 1 & 0 \\
      0 & 0 & 1
    \end{pmatrix}
      & otherwise,
  \end{cases*}
\label{eq:mobility-matrix}
\end{equation}
where $s$ and $v$ are parameters that allows us to model the effects of slip at
the surface.
$m^c_\i$ is the local mobility coefficient for species $c$. For the liquid, we set $m^l_\i$ to be a constant, $m_l$. However, following Ref.~\citenum{robbins2011}, for the nanoparticles we set
\begin{equation}
  m^n_\i = \frac{m_n}2 \bigl( \tanh ( 8\rho^l_\i - 4 ) + 1 \bigr).
  \label{eq:mob_nano}
\end{equation}
This is a smooth function that is $\approx m_n$ when the solvent density is
high and $\approx 0$ when the solvent density is low. This reflects the
fact that the origin of the nanoparticle motion is due to the Brownian
motion from being suspended in the solvent and so the nanoparticles
should be immobile when there is no solvent liquid surrounding
them~\cite{chalmers2017,robbins2011}. It also prevents the nanoparticles
from moving around once the liquid has evaporated. Henceforth we set
$m_n = m_l = 1$, so that all times are given in terms of the Brownian
timescale $\tau_B = \sigma^2/m_nk_BT$.

The parameter $s$ in Eq.~\eqref{eq:mobility-matrix} determines the fluid
mobility parallel to the surface in the first layer of lattice sites
($j=1$). The parameter $v$ controls the mobility
from the first to the second layer, in the direction perpendicular to
the surface. If there is slip, then $s=v=1$. When there is no-slip or
partial-slip then $s=0$ and $v \ll 1$. As we show below, the
receding fluid contact angles are determined by the value of $v$.
{Note that if $s=0$ and $v\sim {\cal O}(1)$, then slip still occurs and the results are almost indistinguishable from results with $s=1$, due to the liquid being able to diffuse from one lattice site on the surface to a neighbouring one indirectly via the layer of lattice sites above at a speed that is faster than the speed of the overall contact line motion.}

We solve this system numerically on a lattice with a finite time step. The divergence and gradients in Eqs.\ \eqref{eq:ddftl} and \eqref{eq:ddftn} are performed using nearest neighbour finite
difference approximations. Care in how these are done is needed to prevent numerical instabilities. We use a mix of
forward and backward finite differences. At a particular time step, if
a forward finite difference for the gradient is used then a backward
finite difference is used for the divergence. As time precedes the
direction of the spatial finite difference is alternated to prevent a
directional bias that can lead to droplets drifting across the surface.
More detail about the finite difference integration scheme is
given in the Appendix.

\section{Results for evaporating droplets}\label{evaporation}

\subsection{Influence of slip at the surface}

First, we discuss the behaviour of evaporating liquid ridges that
do not contain any nanoparticles.
We assume that the fluid density profile only varies in one of the
directions parallel to the surface in order to simplify the numerics.
However, we expect the results to be similar to those one would obtain for a droplet that is initially circularly symmetric with diameter equal to the width of the liquid ridge, so henceforth we refer to them as `droplets'.
Evaporation is simulated by fixing the liquid density of
the lattice sites at the top of the simulation box to a very low value
($10^{-8}$). This emulates an open system with a mechanism, (e.g.\ 
air-flow over the top of a container) for taking the solvent vapour out
of the system. The evaporation rate of a droplet depends very sensitively
on the distance from the top layer of lattice sites from which the liquid is
removed down to the top of the droplet. If the distance is
small the droplet evaporates relatively quickly. On the other hand, if the
distance to the top of the container is large, the droplet evaporation is
slow. Here, we set the system size to be
three times the height of the initial droplet. This sets the overall timescale
for evaporation: it is determined by the time it takes for the vapour to diffuse the height of the system.

\begin{figure}[t]
\includegraphics{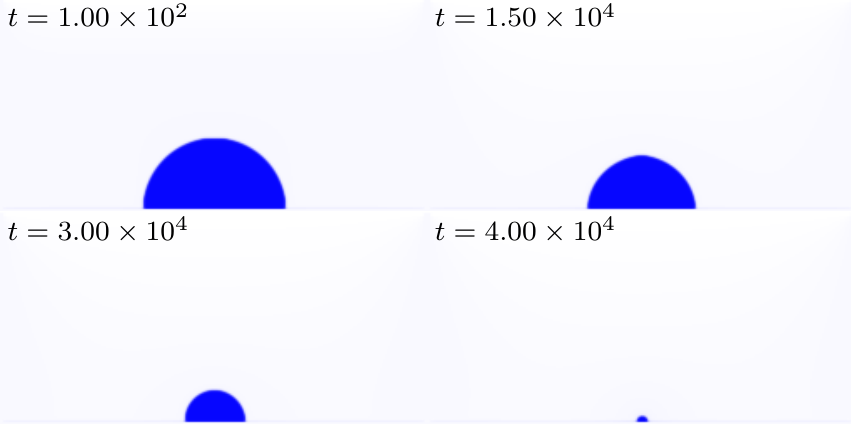}
\caption{%
  Snapshots of a droplet evaporating from a surface calculated using
  DDFT with $s=v=1$ (corresponding to slip at the surface),
  $\e_{wl}/\e_{ll}=0.45$, $k_BT/\e_{ll} = 1.3$ in a $128 \times 64$
  system. The times at which the snapshots occur are given on each
  figure and are in units of the Brownian timescale $\tau_B$.
}
\label{fig:liquid-smooth-evap}
\end{figure}

In Fig.~\ref{fig:liquid-smooth-evap} we display results for a surface with the slip parameters $s=v=1$, for the evaporation of a droplet that initially has semicircular cross-section and with equilibrium contact angle $\approx 90\degree$. As it evaporates, the droplet 
retains its semicircular shape and has a receding contact angle that remains almost equal to the equilibrium contact angle. This is a consequence of the smoothness of the surface on which the droplet is sitting and the slip at the surface, since $s=v=1$.
However, most observed evaporating droplets have (at least initially) a pinned contact line. This is due to the fact that almost
all real surfaces are rough, at least on the microscopic scale.

\begin{figure}
\includegraphics{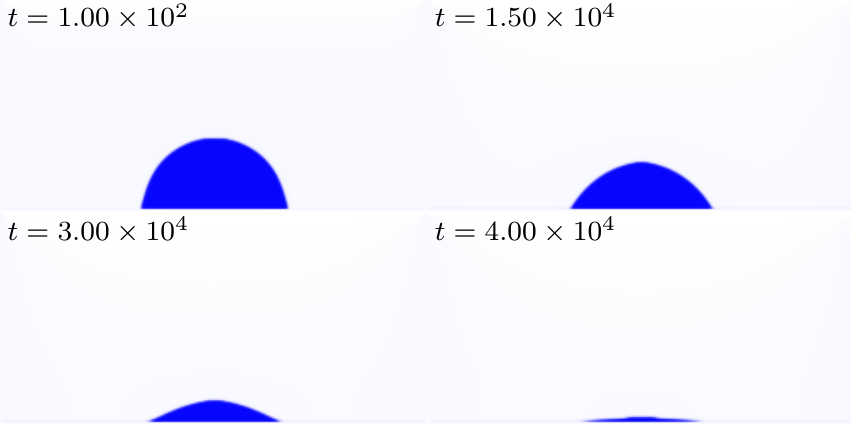}
\caption{%
  Snapshots of a droplet evaporating from a surface
  with $s=0$, $v=0.001$ (corresponding to no-slip at the surface and a
  small receding contact angle), $\e_{wl}/\e_{ll}=0.45$, $k_BT/\e_{ll}
  = 1.3$ in a $128 \times 64$ system. The times are given on each snapshot.
}
\label{fig:liquid-evap}
\end{figure}

\begin{figure}
\includegraphics{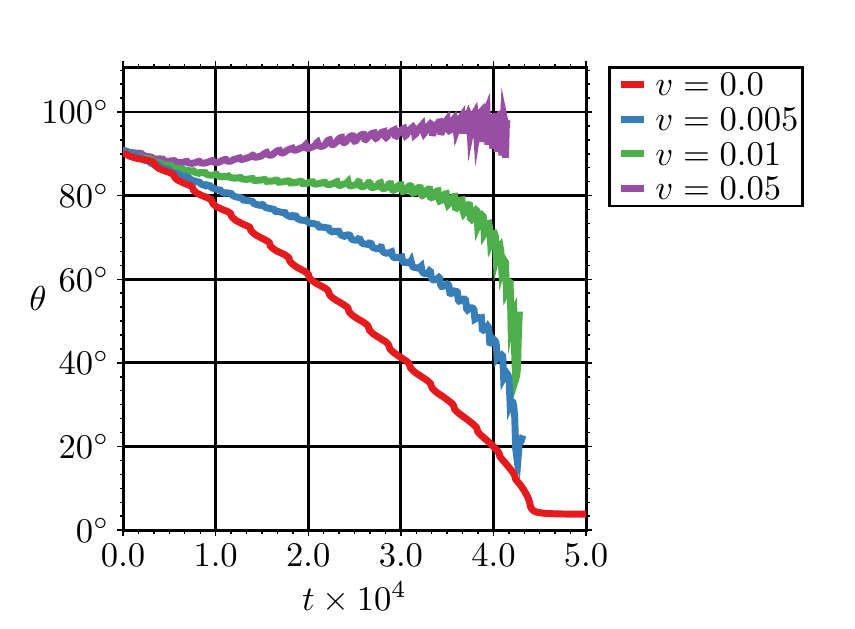}
\caption{%
  The contact angle as a function of time for slip parameters $s=0$ and
  various values of $v$, as given in the key. The line stops when the
  droplet has evaporated (in the $v=0$ case the droplet never fully evaporates  since for this case the $j=1$ layer directly above the surface is completely immobile). The value $v=0.05$ gives very similar results to $v=1$, i.e.\
  for $v \geq 0.05$ the behaviour is independent of $v$.
}
\label{fig:receding-angles}
\end{figure}

By setting the slip parameter $s=0$ we prevent any density exchange
between neighbouring lattice sites directly above the surface, i.e.\ those in the $j=1$ layer. The density in these lattice sites can only vary by exchanging mass with the $j=2$ lattice sites above them. The rate at which this occurs is set by the parameter $v$. Fig.~\ref{fig:liquid-evap} shows snapshots of a droplet evaporating from a surface when the slip parameters $s=0$ and $v=0.001$. We see that the effect of this decreased mobility near the surface is to
pin the contact line, modelling the effect of surface roughness. The
contact lines stay almost stationary at the beginning of the
simulation until the droplet reaches the receding contact angle (in the case in Fig.~\ref{fig:liquid-evap}, this is rather small in value). At this stage the droplet continues evaporating with a moving contact line until it
has completely evaporated. We calculate the contact angle
over time using the circle-fitting method of Refs.\ \citenum{chalmers2017, taubin}. Results from doing this are
shown in Fig.~\ref{fig:receding-angles}, for various values of $v$. We see that when $s=0$ the selected value of $v$
specifies the receding contact angle. Note that there are small oscillations over time in the value of the receding contact angle created by the {way the gas-liquid interface at the} top of the droplet {moves in a discrete manner} from one layer of lattice sites to another {(this can also be seen in Fig.~\ref{fig:liquid-evap} where the evaporating sessile droplet has a small deviation from a spherical cap shape at $t=3.0\times10^4$)}. This artefact of the lattice can also be seen in the underlying binding potential, as discussed in Ref.\ \citenum{hughes2015}. Note also that in the final stages the droplet becomes very small so the circle fitting becomes less and less accurate until eventually it becomes ill-defined.

\subsection{Evaporating droplets of nanoparticle suspension}

Now we consider the evaporation of droplets containing nanoparticles.
Since the vapour density near the top of the simulation box is always low,
due to Eq.\ \eqref{eq:mob_nano} there
is a very low probability of nanoparticles entering the vapour and
escaping the system. The method for incorporating the effects of
surface roughness (i.e.\ no-slip) via Eq.~\eqref{eq:mobility-matrix} is
also applied to the nanoparticles,
using the same values of $s$ and $v$. We assume that the initial density
of the nanoparticles in the liquid droplet is $\rho_n=0.05\rho_l$, i.e.\
the initial concentration $\rho_n / (\rho_l+\rho_n) \approx 0.048$.

\begin{figure*}[t]
\includegraphics[width=\textwidth]{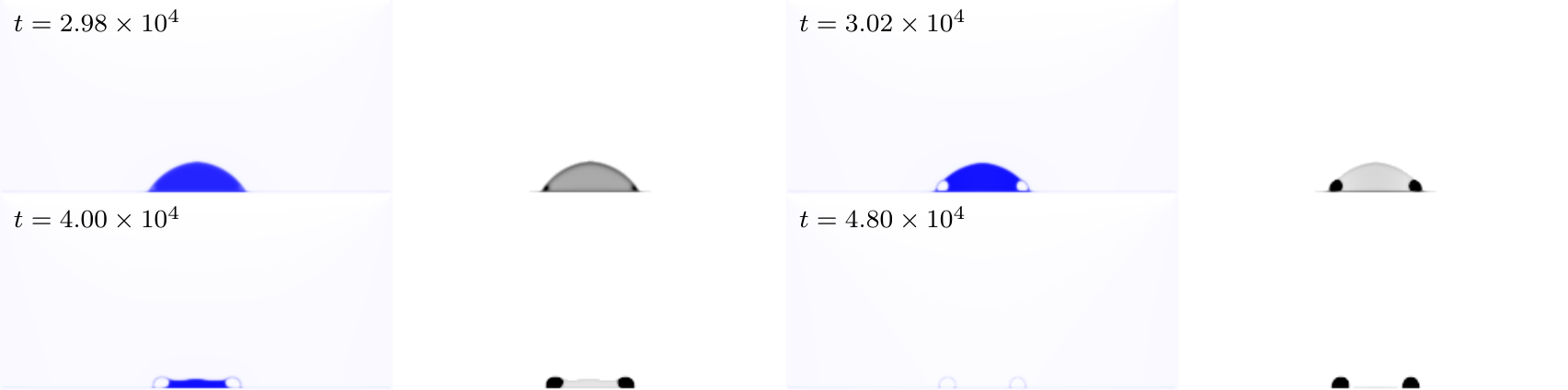}

\includegraphics[width=0.4\textwidth]{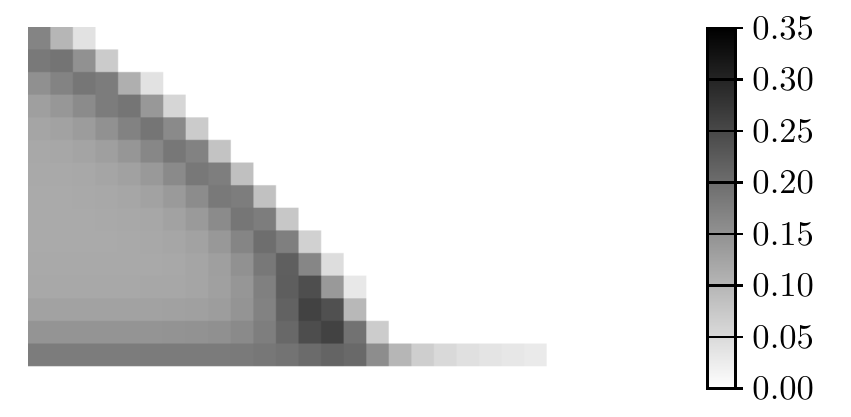}

\caption{%
Top: Snapshots during the evaporation of a droplet of nanoparticle suspension from a surface with $\e_{wl}/\e_{ll} = 0.8$,
  $\e_{ln}/\e_{ll} = 0.75$, $\e_{nn}/\e_{ll} = 1.5$ and $k_BT/\e_{ll} =
  1.3$, $s=0$ and $v=0.005$ in a $256\times128$ system. In each pair of density profiles, the solvent is on the left and the nanoparticle density profile is on the right. {Bottom: Magnification in the contact line region of the nanoparticle density profile for the time $t=2.98\times 10^4$.}
}
\label{fig:nano-noslip}
\end{figure*}

Fig.~\ref{fig:nano-noslip} shows snapshots of a droplet of nanoparticle
suspension evaporating from a surface with almost no-slip ($s=0$, $v=0.005$). As evaporation proceeds the local density of the nanoparticles at the gas-liquid interface builds up to form a crust on the surface of the droplet, with the nanoparticle density at the interface becoming approximately twice the density inside the droplet. {This buildup is even more marked at the contact line, due to the fact that it is very slowly receding which leads to the front gathering even more nanoparticles.} By the time $t=2.98\times 10^4$ the buildup of nanoparticles at the contact line is clearly visible, as can be seen in {the lower panel magnification in Fig.~\ref{fig:nano-noslip}}. This is the start of significant change in the nanoparticle density distribution. A short time thereafter $t=3.00\times 10^4$, the majority of the nanoparticles have collected together at the contact line. This occurs due to phase separation within the droplet, between a nanoparticle rich phase (with a low density of the solvent) and a nanoparticle poor phase (with a high density of the solvent) -- i.e.\ a liquid-liquid phase separation. The phase separation is driven by the fact that as the solvent evaporates, the concentration of the nanoparticles within the droplet slowly rises. The separation is triggered when the nanoparticle density reaches the value where the system becomes unstable to demixing. The phase separation occurs relatively rapidly, depositing the nanoparticles in two piles at the contact lines. By the time $t\approx4.8 \times 10^4$ the bulk of the solvent has evaporated and the majority of the nanoparticles are at the contact lines. Beyond this time, the density profiles no longer change in time.

Recall that we have assumed that the system is invariant in one direction, i.e.\ an evaporating liquid ridge. Thus, the deposited nanoparticles correspond to two parallel lines of nanoparticles deposited where the edges of the liquid ridge was initially located. However, if the droplet had initially been circular, then the nanoparticle deposit would correspond to a ring, like a coffee stain. Note, however, that the mechanism just described for the formation of this structure is completely different from that which is normally invoked for coffee ring stain formation, where it is the advective hydrodynamic fluid flow within the droplet from the centre towards the edge, driven by the evaporation, that leads to a pile-up of the suspended particles at the pinned contact line. In the present model there is no advective hydrodynamics and it is for thermodynamic reasons (i.e.\ the phase separation) that the nanoparticles are deposited at the contact line.

{The evaporation induced phase separation observed in Fig.~\ref{fig:nano-noslip} occurs because the nanoparticles are only weakly soluble in the liquid, since $\e_{nl}/\e_{ll}=0.75$ and $\e_{nn}/\e_{ll}=1.5$; for the chemical potential value $\mu_n/\e_{ll}=-8$, the phase diagram is given in Fig.~\ref{fig:liquid-binodal}(d). However, if the nanoparticles are more soluble in the liquid [e.g.\ with $\e_{nl}/\e_{ll}=1.25$ and $\e_{nn}/\e_{ll}=1.5$, corresponding to the phase diagrams in Fig.~\ref{fig:liquid-binodal}(b) and (c)], then we observe no phase separation or line deposition and the nano-particles are deposited uniformly on the surface (not displayed).}

\begin{figure*}
\includegraphics[width=\textwidth]{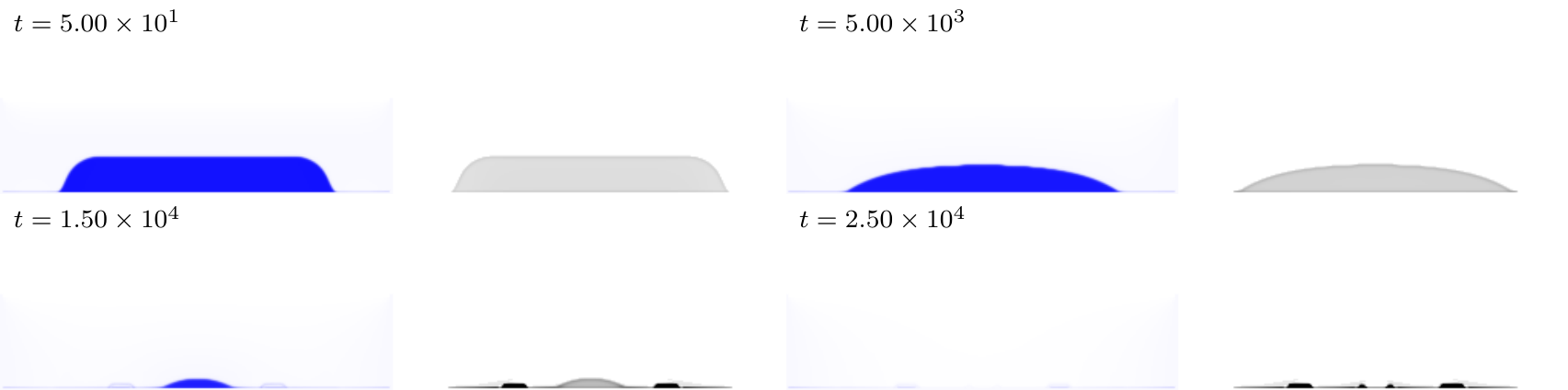}
\caption{%
  Snapshots taken from simulating a droplet of nanoparticle suspension
  evaporating from a surface with $\e_{wl}/\e_{ll}=0.8$,
  $\e_{ln}/\e_{ll} = 0.75$, $\e_{nn}/\e_{ll} = 1.5$, $k_BT/\e_{ll} =
  1.3$ and slip $s=v=1$ in a $128 \times 64$ system.
}
\label{fig:nano-long}
\end{figure*}

In Fig.~\ref{fig:nano-long} we present results for a case where there is slip at the surface (i.e.\ $s=1$, $v=1$), in contrast to the previous case in {Fig.\ \ref{fig:nano-noslip}.} Fig.~\ref{fig:nano-long} displays snapshots from the evaporation of a wider droplet that is initially pancake-like, with a flat top. As the liquid evaporates, a crust of nanoparticles still forms at the gas-liquid interface, but in this case the contact line also recedes, due to the slip at the surface. There is a buildup of nanoparticles at the contact line, which is enhanced compared to the case in {Fig.\ \ref{fig:nano-noslip}}, due to the fact that the contact line is receding. As it recedes, the demixing transition is triggered, so there is a deposition of the nanoparticles partway between the initial location of the contact line and the centre of the droplet. Furthermore, not all of the nanoparticles are deposited at this stage, due to the larger size of the droplet. As evaporation continues, the contact lines de-pin from the nanoparticle deposits and further recede. There is therefore again an increase in the concentration of the nanoparticles in the droplet and also a buildup at the contact line until again the phase transition is triggered, leading to a second deposition of nanoparticles closer to the centre. These deposits are somewhat smaller than the first.

We believe that for different parameter values and for even larger droplets, this process could lead to the formation of a deposit consisting of a greater number of concentric rings and perhaps even of a periodic nanoparticle deposition. This would appear similar to the periodic nanoparticle deposition process described in Refs.\ \citenum{thiele2014patterned, thiele2009, frastia2011, frastia2012} and the (experimental) references therein. However, the mechanism here is entirely different: it is due to the thermodynamics of phase separation, rather than due to advective fluid-dynamics.

\begin{figure*}
\includegraphics{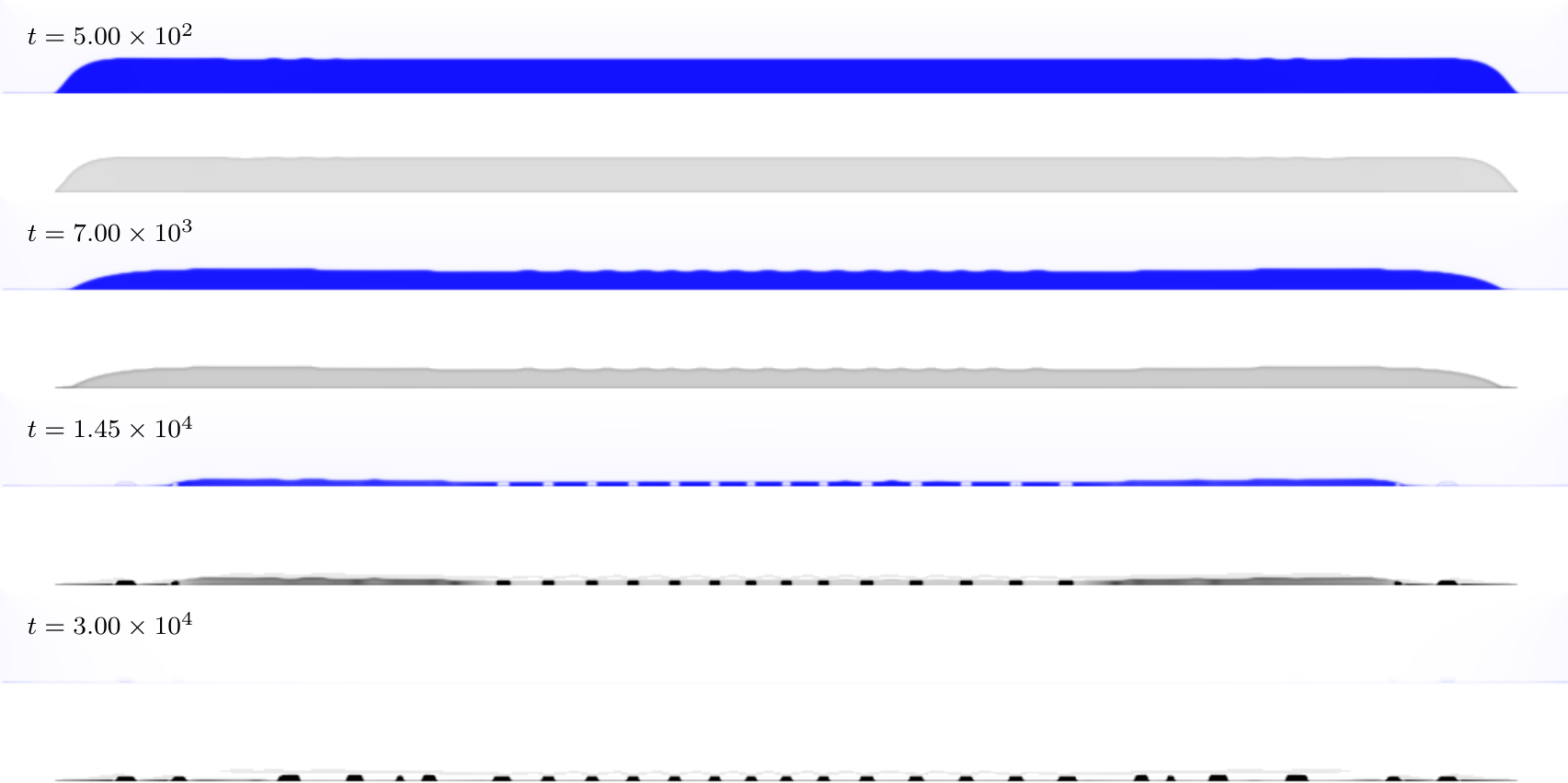}
\caption{%
  Snapshots taken from simulating a pancake-like droplet of
  nanoparticle suspension
  evaporating from a surface with $\varepsilon_{wl}=0.8$,
  $\varepsilon_{ln} = 0.75$, $\varepsilon_{nn} = 1.5$ and $k_BT/\e_{ll} = 1.3$
  in a $1024 \times 64$ system.
}
\label{fig:nano-v-long}
\end{figure*}

For the parameter values used here, when we keep the initial height of the droplet the same, but make the width greater, we see in Fig.~\ref{fig:nano-v-long} something different: At the receding contact line, we still see the deposition just described. However, in the middle of this wide pancake-like droplet, when the concentration of the nanoparticles reaches a high enough level due to the evaporation, we see spinodal demixing occurring in the middle of the film. This has a characteristic wavelength and so leads to a periodic array of nanoparticle deposits on the surface. The characteristic wavelength is also seen in the small amplitude periodic modulation in the thickness of the liquid film, that is a precursor to the demixing. The nanoparticle deposits occur where the troughs are located. In this situation the film is so thin that the surface and what is left of the bulk of the film are strongly coupled. The coupling of demixing within the film to the film height has been observed e.g.\ in films of polymer blends \cite{thomas2010wetting}. For a detailed discussion of demixing in thin liquid films and how this may couple to the film hight profile, see Ref.~\citenum{bribesh2012decomposition}. There may be regimes where this leads to demixing induced front instabilities \cite{vancea2011pattern, yerushalmi1999alternative}.

\section{Concluding remarks}\label{conclusion}

We have described a DDFT model for the evaporation of droplets of a nanoparticle suspension from surfaces. We have shown that the model can include the effects of (no)slip at the surface, nanoparticle crust formation and nanoparticle aggregation which leads to the deposition of ring deposits and other more complex structures.

A particularly striking result of the present work is the observation that the coffee stain effect can still arise in a system with no advective hydrodynamics to carry the suspended nanoparticles to the contact line. Here we show that the thermodynamics of aggregation and phase separation can also lead to the formation of ring stains. Furthermore, it can also lead to the formation of multiple rings. Further work is required to see if there is a regime within the present model where periodic line deposition can be observed, as in certain hydrodynamic models \cite{frastia2011, frastia2012}. {Note that there are numerous experimental systems where periodic deposition is observed -- see the comprehensive review in Ref.~\citenum{thiele2014patterned}.} We believe there probably is a regime {in the present model} where periodic deposition occurs, however for the parameter values and system sizes we have so far explored, we have not observed this.

{It is straightforward to extend the current model to include additional species of either liquid or nanoparticles. Each new species introduces an additional density profile for each and the Helmholtz free energy in Eq.\ \eqref{eq:helmholtz} generalises in a straight-forward manner. Lattice DFTs have also been developed to describe elongated or other shaped particles -- see for example the recent work in Ref.\ \citenum{mortazavifar2017phase}. To describe the non-equilibrium dynamics of such systems, one should simply insert the given approximation for the equilibrium Helmholtz free energy into the generalisation of the DDFT equations \eqref{eq:ddftl} and \eqref{eq:ddftn} that also include additional terms related to rotational diffusion.\cite{wittkowski2011dynamical}}

The main direction where the present approach should be extended is to incorporate the advective hydrodynamics of the solvent liquid. The DDFT is certainly adequate for describing the dynamics of the nanoparticles. However, for the solvent dynamics, especially for situations such as sliding droplets on inclined planes (see e.g.\ Ref.~\citenum{thiele2003front}) {or any situation where the fluid is strongly driven or when the flow contains swirls, eddies, etc., then the present approach is unlikely to be} adequate. That said, since the theory is based on a free energy functional incorporating the correct thermodynamics, it should remain qualitatively correct, at least for when the system is not driven too far from equilibrium. {This is because when the system evolution can be well-approximated by a conserved dynamics that at all times seeks to decrease the free energy, distinguishing between diffusive or advective motion generally only changes the rate at which the system evolves through the free energy landscape, but not the pathway,\cite{honisch2015instabilities, yin2017films} and therefore the dynamics is qualitatively the same in both cases.}

Indeed, this is one of the great advantages of using DFT and DDFT as the theoretical foundation: these base the theory on a free energy function(al) -- in the present case, Eq.\ \eqref{eq:helmholtz} -- so as a result, this gives easy access to important thermodynamic quantities such as the pressure, surface tension and equilibrium contact angle. Obtaining these from MC simulations or other approaches is generally more complicated and requires more lengthy computations \cite{chalmers2017}.

\section*{Acknowledgements}

The authors thank Adam Brunton, David Sibley and Dmitri Tseluiko for useful discussions on our work.

\section{Appendix}

We integrate the coupled Eqs.~\eqref{eq:ddftl} and~\eqref{eq:ddftn}
forward in time using the Euler algorithm:
\begin{equation}
  \rho^c_\i(t + \Delta t)
    = \rho^c_\i(t)
    + \frac{\partial\rho^c_\i}{\partial t} \Delta t,
\label{eq:euler}
\end{equation}
where $c=l,n$ and we replace the terms $\partial \rho^c_\i /
\partial t$ by the respective expressions in either the right hand side of
Eq.~\eqref{eq:ddftl} or Eq.~\eqref{eq:ddftn}. We use the value $\Delta t
= 10^{-5}$.

To evaluate the spatial finite differences in the direction parallel to the surface, we do the following:
Let $d$ be the direction in which we take the derivative. $d$
alternates between $1$ and $-1$ from one time step to the next to
prevent any direction bias. The following quantity
\def\dfdc{\frac{\partial F}{\partial \rho^c}}
\begin{equation}
\rho^c\nabla \left. \frac{\partial F}{\partial \rho^c} \right|_{(i,j)} =
  \left( I^c_{(i,j)}, J^c_{(i,j)} \right),
\label{eq:finite-ij}
\end{equation}
which occurs in the right hand sides of
Eqs.~\eqref{eq:ddftl}~and~\eqref{eq:ddftn}, is evaluated as
\begin{align}
I^c_{(i,j)} &=
  \frac{\rho^c_{(i,j)} + \rho^c_{(i+d,j)}} 2
  \left( \dfdc\evalAt{(i+d,j)} - \dfdc\evalAt{(i,j)} \right),
\label{eq:finite-i}
\\
J^c_{(i,j)} &=
  \frac{\rho^c_{(i,j)} + \rho^c_{(i,j+1)}} 2
  \left( \dfdc\evalAt{(i,j+1)} - \dfdc\evalAt{(i,j)} \right).
\label{eq:finite-j}
\end{align}

Note that we have assumed that for all time the densities remain invariant
in the direction indexed by $k$, where $\i=(i,j,k)$. If this is not the
case, then there is an additional component of the same form as
Eq.~\eqref{eq:finite-i}.

Multiplying Eq.~\eqref{eq:finite-ij} by the respective mobility matrix
from Eq.~\eqref{eq:mobility-matrix} we obtain

\begin{equation}
  (X^c_\i,Y^c_\i) =
  M^c_\i\rho^c_\i\nabla \frac{\partial F}{\partial \rho^c_\i}.
\end{equation}
Now when we take the divergence we use the opposite direction $d$:
\begin{equation}
\nabla \cdot (X^c_\i,Y^c_\i) =
  X^c|_{(i,j)} - X^c|_{(i-d,j)} +
  Y^c|_{(i,j)} - Y^c|_{(i,j-1)}.
\end{equation}

This is the expression we use for $\partial\rho^c_\i / \partial t$ in
Eq.~\eqref{eq:euler}.

\providecommand{\latin}[1]{#1}
\makeatletter
\providecommand{\doi}
  {\begingroup\let\do\@makeother\dospecials
  \catcode`\{=1 \catcode`\}=2 \doi@aux}
\providecommand{\doi@aux}[1]{\endgroup\texttt{#1}}
\makeatother
\providecommand*\mcitethebibliography{\thebibliography}
\csname @ifundefined\endcsname{endmcitethebibliography}
  {\let\endmcitethebibliography\endthebibliography}{}

\end{document}